\documentstyle[aps,preprint,graphicx]{revtex}
\begin{document}
\draft
%\twocolumn[\hsize\textwidth\columnwidth\hsize\csname@twocolumnfalse\endcsname
\title{Real event detection  and the treatment of congestive heart failure: an efficient technique to help cardiologists to make crucial decisions}
\author{P. Allegrini$^{1}$, R. Balocchi$^2$, S. Chillemi$^3$, P. Grigolini$^{3,4,5}$, P. Hamilton$^{6}$, R. Maestri$^{7}$,
L. Palatella$^{5}$, G. Raffaelli$^{8}$}
\address{$^{1}$Istituto di Linguistica Computazionale del Consiglio Nazionale delle
Ricerche, \\
Area della Ricerca di Pisa-S. Cataldo, Via Moruzzi 1, 
56124, Ghezzano-Pisa, Italy }
\address{$^{2}$Istituto di Fisiologia Clinica del Consiglio Nazionale delle
Ricerche, \\
Area della Ricerca di Pisa-S. Cataldo, Via Moruzzi 1, 
56124, Ghezzano-Pisa, Italy }
\address{$^{3}$Istituto di Biofisica del Consiglio Nazionale delle
Ricerche, \\
Area della Ricerca di Pisa-S. Cataldo, Via Moruzzi 1, 
56124, Ghezzano-Pisa, Italy }
\address{$^{4}$Center for Nonlinear Science, University of North Texas,
P.O. Box 311427, Denton, Texas, 76203-1427}
\address{$^{5}$Dipartimento di Fisica dell'Universit\`{a} di Pisa and INFM 
Via Buonarroti 2, 56127 Pisa, Italy }
\address{$^{6}$Center for Nonlinear Science, Texas Woman's University, 
P.O. Box 425498, Denton, Texas 76204}
\address{$^7$ 
Fondazione S. Maugeri, IRCCS, 
Istituto Scientifico di Montescano,
Via per Montescano,
27040 Montescano (PV), Italy.
% Maugeri Foundation, IRCCS, Rehabilitation Institute of Montescano, Pavia, Italy
}
\address{$^8$International School for Advanced Studies and INFM Unit,
via Beirut 2-4, 34014 Trieste, Italy}
\maketitle
\date{\today}
\maketitle

\begin{abstract}
Using a method of entropic analysis of time series we establish the
correlation between heartbeat long-range memory and
 mortality risk in patients with congestive heart failure. 
\noindent
%02.50.Ey,05.20-y.05.40.Fb,05.70-a}
\end{abstract}
%\vskip1pc]
%\narrowtext
%\pacs{02.50.Ey,05.20-y.05.40.Fb,05.70-a}

This paper deals with an advanced aspect of statistical mechanics whose
recent demonstration \cite{memorybeyondmemory} is proven here to
afford an efficient criterion to deal with the mortality
risk of patients with Congestive Heart Failure (CHF). We define the
condition corresponding to the mean position of healthy subjects in a
plane, called, as we shall see, physiological plane. We show
that the distance of a CHF
patient from this optimal condition correlates with mortality, 
and that the CHF patients very close to it survived.  
The importance of this criterion
does not need to be defended. We think that the demonstration of this
important property is of interest for both researchers at the
frontier of statistical mechanics and cardiologists. The first step
of our approach is based on observing a sequence of numbers
$\{T_{i}\}$, denoting the distance between two nearest neighbor pulses
of a given electrocardiogram (ECG). One way to denote the time between
two pulses is to measure the time elapsed betwen two adjacent R waves
in the recorded electrical signal. This time is referred to as the RR
interval, and the resulting sequence is consequently denoted as RR
sequence.

%We considered 29 moderated to severe CHF male patients, from a study
%base of 320 subjects consecutively referred to the Division of
%Cardiology between 1992 and 1999 for evaluation and therapy (follow-up
%time: 16+-12 months, median: 12). All subjects had a 24-hour Holter
%recording at baseline. 
We considered for this study 13 male CHF patients, from a study base
of 320 subjects, who experienced cardiac death during a follow-up of
26 months (average 19 months, median 22 months). 
Inclusion criteria were absence of
pulmonary or neurological disease, absence of acute myocardial
infarction or cardiac surgery within the previous 6 months, absence of
any other disease limiting survival, stable therapy for at least 2
weeks and good quality 24-hour Holter recordings, with an ectopy rate
less than 5\%. A comparable number of control subjects (16
patients), 
matching for age, sex, NYHA class (a functional and therapeutic 
classification for prescription of physical activity for cardiac patients) 
and etiology, was then selected. 
These latter patients did not experience cardiac death after follow-up.
All patients had
a 24-hour Holter recording at baseline, together with standard
functional evaluation including measurement of left ventricular
ejection fraction (LVEF), peak VO2 and Sodium (Na). 
Finally, RR series for 10 healthy subjects were
taken from the {\em NOnLinear TIme Series AnaLysIS}
(NOLTISALIS) archive.
This latter data set is the result of the collaboration of several
interdisciplinary Italian research centers.
Experienced analysts edited
these Holter recordings, manually correcting 
interbeat times due to ectopic beats. 
This editing work yielded the RR sequence,
from which we generated the time series $\{T_i\}$.

%Holter recordings were manually edited by
%experienced analysts and corresponding annotated RR time series 
%$\{T_i\}$ derived. Before the follow-up date, 13 patients died
%or had to receive heart transplant to survive. This group is
%referred to as ``dead patients''.

The meaning of a given value $T_{i}$, as earlier stated, is the time
distance between the $i-th$ and the $(i+1)-th$ pulse.
The sequence $\{T_i\}$ can be studied as a new time-series, with
$i$ playing the role of ``time''.
Moreover, the value $T_{i}$,
expressed as a function of $i$ with $i \gg 1$, can be thought of as a
function $T(t)$, namely, as a function of a continuous time variable $t$.

The real curve $T(t)$ looks erratic and disordered. However, our
method of analysis shows that it is quite different from a 
random process. For the reader to get an intuitive understanding of
this attractive but perplexing conclusion, let us describe first an
ideal model, with extended memory (EM), for the time evolution of
$T(t)$. First of all, we assume that for a given time $\tau_{em}$, the
curve $T(t)$ keeps a given slope $\alpha$, then it abruptly gets a new
slope, $\alpha^{\prime}$, for an interval of time
$\tau_{em}^{\prime}$, after which a new abrupt transition to a new
slope $\alpha^{\prime \prime}$ takes place, for a time
$\tau_{em}^{\prime \prime}$, and so on. It is evident that the
resulting $T(t)$ has the form of a zig-zag curve. We shall refer to
the individual straight line intervals of this curve as \emph{laminar
regions}. Any laminar region is associated with its own
$\tau_{em}$. Then we introduce the extended memory property. This is
done by assuming for the waiting time distribution $\psi(\tau_{em}) $
the following inverse power law form

\begin{equation}\label{crucialform}\psi(\tau_{em}) = (\nu -1)
\frac{\left[\langle \tau_{em} \rangle (\nu-1) \right]^{\nu-1}}
{ \left[\langle \tau_{em} \rangle (\nu-1) + \tau_{em}\right]^{\nu}},
\end{equation}
with $2 < \nu < 3$, where $\langle \tau_{em} \rangle$ is the
average waiting time. This means that if the EM model is directly
observable, we can derive from $T(t)$ the sequence $\{\tau_{em}
(j)\}$, where the discrete index $j$ denotes the time order of a given
laminar region.  

It is interesting to notice that this dynamic process is essentially
equivalent to the strong anomalous diffusion model recently proposed
by the authors of Refs.\cite{vulpiani1,vulpiani2} to explain the
effects of a ballistic mechanism in the Rayleigh-B\'{e}nard
convection. 
%This model, in turn, is nothing but the a generalization
%of that proposed years ago for a dynamic derivation of L\'{e}vy
%statistics \cite{allegro} that corresponds to assuming that the slope
%$\alpha$ can only have two values, one positive and one
%negative\cite{note}.  
This model, in turn, is nothing but a generalization of the dynamic
approach to L\'{e}vy statistics proposed years ago by the authors of 
Ref. \cite{allegro}.
In fact, the model of Ref. \cite{allegro} is recovered from the model we are 
adopting here, by assuming that the slope $\alpha$ has only two 
values, of equal intensity and opposite sign \cite{note}.
With this equivalence in mind, we adopt the
specific walking prescription proposed by Ref.\cite{giacomo}.  We
consider a given time $t$ and we evaluate the number of laminar
regions that have been completed within this time. Let us call 
this number $n$. Then the trajectory is

\begin{equation}
y(t) = n W.\label{trajectory}
\end{equation} 
This means that the random walker makes a jump ahead, by the same
quantity $W$, at the end of any laminar region. The quantity $W$ is
arbitrary and in the following we assume $W=1$. Then, according again
to the prescriptions of Ref. \cite{giacomo}, we create the
trajectories

\begin{equation}
\label{manytrajectoriesfromone}
x(l) = y(t+l) - y(t).
\end{equation}
It is evident that we can move the index $t$ from $t=0$ to $L-l$,
where $L$ is the total time length of the sequence under study.  This
makes it possible for us to evaluate the probability distribution
$p(x,l)$, which at $l=0$ is a delta of Dirac located at $x = 0$,
broadening upon time increase. Note that an important step of our
approach rests on the determination of the Shannon entropy of this
distribution, namely

\begin{equation}
S(l) = - \int_{\infty}^{\infty} dx p(x,l) \log p(x,l).
\label{diffusionentropy}\end{equation}
This is the reason why this technique of analysis has been termed
Diffusion Entropy (DE) method \cite{nicola}. According to the theory
of Ref.\cite{giacomo} we immediately reach the conclusion that
$p(x,l)$ fulfils the scaling property
\begin{equation}\label{scaling}
p(x,l) = (1/l^{\delta} )
F(x/l^{\delta}),
\end{equation}
with $\delta$ being the scaling
parameter, which is related to $\nu$ by

\begin{equation}\label{deltaasfunctionofmu}
\delta = 1/(\nu -1).
\end{equation}
In this specific case, it is straightforward to prove, by plugging
$p(x,l)$ of Eq. (\ref{scaling}) into Eq.(\ref{diffusionentropy}),
that\begin{equation}S(l) = A + \delta \log
(l).\label{logarthmicdependence}\end{equation} This means that the DE
should yield for $S(t)$, expressed in a linear-log plot, a straight
line, whose slope is the searched value of the scaling parameter
$\delta$.

This way of proceeding is impossible in practice, because the real
$T(t)$ curve significantly departs from the zig-zag form of the EM
model.  We make the conjecture that the departure from this ideal
condition is caused by the fact that the actual signal $T(t)$ is the
superposition of the EM model signal and a much stronger, but totally
random component. This makes it impossible for us to directly
evaluate $\psi(\tau_{em})$. We proceed as follows. Let us represent the $T(t)$
time evolution in a $(T,t)$ plane, with the ordinate referring to $T$
and the abscissa to $t$. The ordinate axis is divided into cells of
equal size, called $s$. This means that we divide the $(T,t)$ plane
into strips of size $s$, and that in the ideal case of constant
frequency the trajectory $T(t)$ would move forever remaining within
the same strip. Actually, transitions from one
strip to the other occur frequently. 
We call these transitions markers. These markers
might have quite different origins. The majority of the markers are
determined by the short-time noise. Many other markers correspond to the
hidden laminar regions of the underlying EM model; 
we call these markers ``pseudoevents''. 
As in Ref. \cite{memorybeyondmemory}, we indicate with the term pseudoevent a
marker that does not correspond to an unpredictable transition, but it
is a consequence of the division of the $(T,t)$ plans into
strips. Only a very small number of markers coincide with 
the turning points of the EM signal, or are sufficiently close to them. 
We call these markers \emph{real events}.

Now we have to explain why the DE is sensitive only to the real
events. The time distance between the $j$-th and the $(j+1)$-th marker,
defines the time $\tau_{exp}(i)$ of the experimental sequence
$\{\tau_{exp}(i)\}$. It is evident that even in the case where $T(t)$
were an exact realization of the EM model, the waiting time
distribution $\psi(\tau)$ might turn out to be totally different from
$\psi(\tau_{em})$.  In fact, if $s$ is very small the same laminar
region is divided into many smaller time intervals, with the same
length. These are pseudoevents. It is evident that these pseudovents
do not contribute to the spreading of the distribution $p(x,l)$, and
consequently do not contribute to the entropy increase. This means
that the DE method is insensitive to pseudoevents.

What about the short-time random events? They, in principle,
contribute the entropy increase, and consequently could affect the
determination of the crucial parameter $\delta$. We can show that they
do not.  Notice that in the presence of the short-time random
component the actual signal $\tau_{exp}(t)$ is given by

\begin{equation}
\tau_{exp}(t) = a \tau_{st}(t) + b R(t),
\label{equation}
\end{equation}
with $a \ll 1$ and $b$ close to $1$.  The first contribution
corresponds to the EM model, and the second is generated by short-time
random fluctuations.  The correlation function of this signal
is
\begin{equation}
\label{experiment}
C_{exp}(t) =  p C_{st}(t) + (1-p) C_{random}(t),
\end{equation}
where $C_{st}(t)$ is an inverse power law relaxation and
$C_{random}(t)$ a relaxation function decaying to zero in one time
step. 
If we take, without loss of generality
$\langle {\tau_{st}}^2\rangle = \langle R^2 \rangle = \langle {\tau_{exp}}^2 \rangle$, implying $ a^2 + b^2 =1$, then we have $p=a^2$ and $(1-p) = b^2$.
How should parameter $p$ be evaluated?  This is easily done
monitoring the experimental correlation function at the first time
step. According to the fact that $C_{random}(t)$ decays to zero in one
step, while $C_{st}(t)$ is much slower, we immediately obtain $p =
C_{exp}(1)$.  Its evaluation is not independent of that of the other
parameter, $ s$. This is so because the experimental evaluation of
$\tau_{exp}(t)$ is dependent on $s$.

However, while $p$ is $s$-dependent, the parameter $\delta$ is
not. The DE method has the surprising capability of yielding a value
for $\delta$ that is independent of $ s$, even in the case when a
strong short-time random component is present, and not only when the
ideal EM model applies. How can it be so?  This is so because the EM
model component yields superdiffusion, while the random component
generates ordinary diffusion. In the asymptotic limit of very large
values of $l$, the superdiffusion component, which is faster than the
ordinary diffusion, becomes predominant, and the DE method detects
again the correct scaling $\delta$ even in the case where $p \ll 1$.

The parameter $p$ is very important, since it defines the statistical
weight of the EM component present in the experimental signal
$T(t)$. However, its dependence on $s$ makes its use questionable.
However, we see that all curves $p(s)$ share the same properties of
getting small values for both small and large values of $s$, with a
clearly defined maximum in between, which is referred to by us as
$\pi$. This maximum is a property independent of $s$. Infact, we note
that almost all the healthy patients get their maximum at 30 ms, while
most of the CHF patients get theirs at 20 ms. The typical dependence
of $p$ on $s$ is illustrated in Fig. 1, where we can see both the case
of 5 healthy subjects with $\pi$ located at 30 ms, and 5 CHF patients,
with $\pi$ located at 20 ms. We think that the parameter $\pi$ is a
reliable measure of the EM component. Consequently, we decided to
represent the physiological conditions of all patients, healthy and
CHF alike, in the $(\delta, \pi)$ that we call \emph{physiological
plane}. The criteria adopted to define the physiological plane make
the resulting ``phase-space'' diagram independent of the coarse-graining
parameter $s$, and the location of any patient in the physiological
plane is an objective property independent of the coarse graining
parameter $s$. A mere visual inspection shows that the healthy and CHF
patients do not mix but in relatively small region. The overlap region
is so small as to make it possible for us to claim that healthy and
CHF patients belong to two distinct regions of the physiological plane
\cite{note2}.

%We define point $A = (\delta_{av},
%\pi_{av})$, which is the average position of healthy patients, and we
%rank the CHF patients in order, according to the distance between
%their position and the point $A$. We find that the first 7 patients
%are alive. We apply the Mann-Withney test\cite{test} to assess the
%probability that this ordering is random . We conclude that the
%probability that this ranking is merely casual is less than
%$3\%$. Taking into account that physicians credit a prediction with a
%proobability of being wrong of less than $5\%$ as a significant
%prediction, we reach the conclusion that our criterion of analysis is
%a reliable way to establish the mortality risk of the CHF patients.

The division of the CHF patients into two groups, dead and alive, make
the result of our analysis still more remarkable.  To show this
important property we proceed as follows. First of all, we define the
center of gravity of the healthy patient, denoted with a white square
in Fig. 2. We call this point optimal condition. Then, for any CHF
patient we measure the Euclidean distance from the optimal condition,
thereby making it possible for us to rank the CHF patients in order,
according to this distance. In other words, the first CHF patient is
the one with minimum distance from the optimal condition. Then we
observe the remaining patients, and we rank as second the one with
minimum distance from the optimal condition, and so on.  We find that
the first 7 patients are alive. The eighth patient is dead, and from
now on the patients are either alive or dead. This suggests that the
closer the patient to the optimal condition the higher the survival
probability.

% fig 1 can be put here

To support in a more rigorous way this important property, we apply
the Mann-Whitney method \cite{test}. This is a non-parametric test,
namely it does not rest on Gaussian distributions for the data.  Let
us count the number of patients dead, $N_{dead}$, and the number of
patients alive, $N_{alive}$. The data of Fig. 2 refer to a case where
$N_{dead} = 13$ and $N_{alive} = 16$. Let us consider the group with
the smaller number of individuals. This means the group of dead
patients. Then, let us evaluate the sum of the ranks of this group
and let us call it $\mu_{exp}$. We note that $\mu_{exp} = 246$.  Under
the hypothesis of no correlation between our parameters and the death
probability, this resulting sum has a probability distribution that,
for more than 8 elements is expected \cite{test} to be approximately a
Gaussian with mean $\mu_T$ and standard deviation $\sigma$ given by

\begin{equation}
\mu_{dead}=N_{dead} \frac{N+1}{2} \:\:\: \sigma=\left ( N_{dead} N_{alive} 
\frac{N+1}{12} \right )^{1/2}. 
\end{equation}
From the data of Fig. 2 we obtain $\mu_{dead}=195$ and $\sigma=22.8$,
while, as earlier noticed, $\mu_{exp}=246$. This means that the
distance of $\mu_{dead}$ from $\mu_{exp}$ is $2.37 \sigma$.  In
practice, we are allowed to rule out the hypothesis of no correlation
between the death of CHF patients and their distance from the optimal
condition: the probability for the distribution of dead and alive
patients of Fig. 2 to be fortuitous, is less than 3\%.

% fig. 2 can be put here

Moreover, it is important to remark that the alive patients corresponding to
points in the physiological plane far from the optimal conditions,
either had a serious pathology, being classified as NYHA class III
(severe physical limitations, they are confortable only at rest) 
and therefore required a heart transplant anyway, or had a
very short follow-up time (less than 6 months). 
Only six alive patients did not belong to either of the above conditions,
but it is remarkable that all six of them occupy positions which 
overlap with the zone of the healthy subjects.
Unfortunately, the small number of sequences available to us at this time
does not allow us to calculate the survival curves, or attempt
any further conclusion.

It is convenient to compare the result of this paper to the literature
in the field of nonlinear or fractal analysis of chardiological data. 
The most recent examples are those of
Refs.\cite{memorybeyondmemory,stanley1,stanley2}. Although based on
different perspectives, multifractality \cite{stanley1,stanley2} and
memory beyond memory\cite{memorybeyondmemory} as a sign of healthy
physiological condition, neither of these two groups could address the
ambitious step of granting physicians a reliable criterion to make
crucial decisions about the CHF patients. 
%It is evident that using the
%statistical method illustrated in this paper it would be possible for
%the physicians to proceed with heart transplantation as soon as
%possible if the CHF patient position in the physiological plane exceeds
%a critical distance from the optimum physiological condition.
These findings, however suggest that $\delta$ and $\pi$ indexes, and
especially the distance from the optimum condition in the
physiological plane, should be considered for inclusion in the
candidate predictors' list of future large-scale prospective studies
for risk stratification of CHF patients.

\emph{Acknowledgements}.  We gratefully acknowledge financial support
from the Army Research Office through Grant DAAD 19-02-0037. One of
the authors (P. A.) acknowledges European Commission POESIA project
(IAP2117/27572) for financial support. L. P. acknowledges
ENEL Research Grant 3000021047 for support. We also thank
S. Maugeri Foundation for supplying the data.

\newpage

\begin{figure}[t]
\vspace{2cm}
\noindent
\hspace{5cm} {\Large \bf a)} \vspace{-2cm}
\begin{center}
\includegraphics[angle=0, width=12cm]{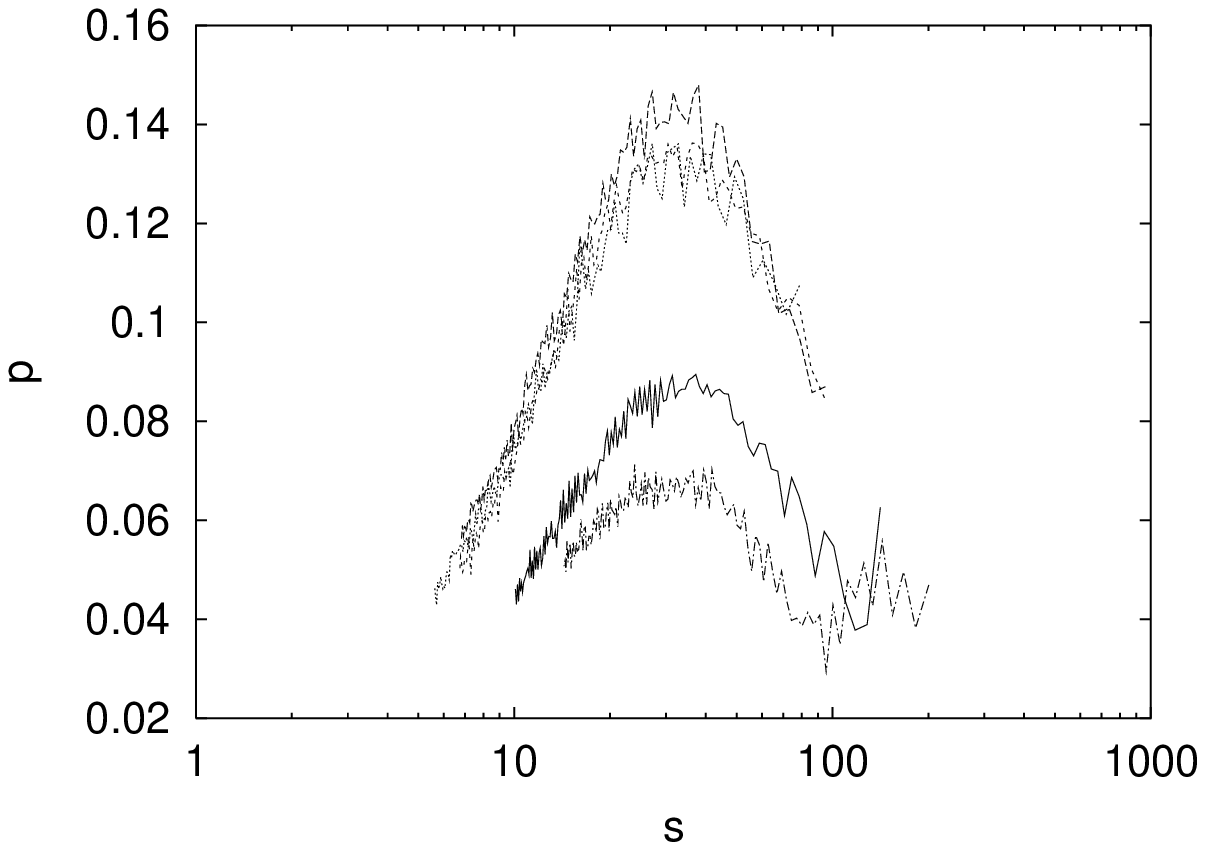}
\end{center}
\vspace{0.5cm}
\hspace{5cm} {\Large \bf b)} \vspace{-2cm}
\begin{center}
\includegraphics[angle=0, width=12cm]{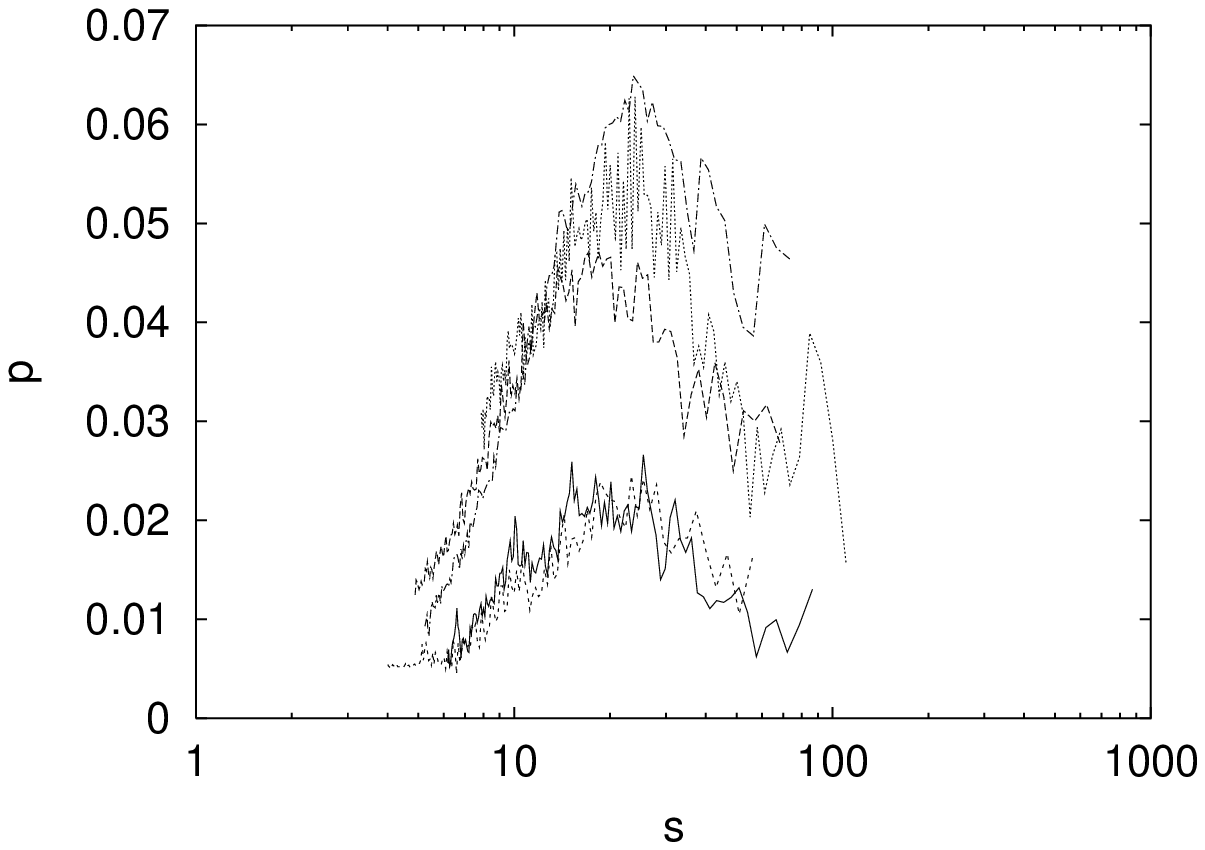}
\end{center}
\caption{\label{fig1} $p$ as a function of $s$ for: a) a group of
healthy subjects; b) a group of CHF patients. For clarity, we have
plotted only 5 subjects for each group.}
\end{figure}

\newpage

\begin{figure}[ht]
\begin{center}
\includegraphics[angle=0, width=12cm]{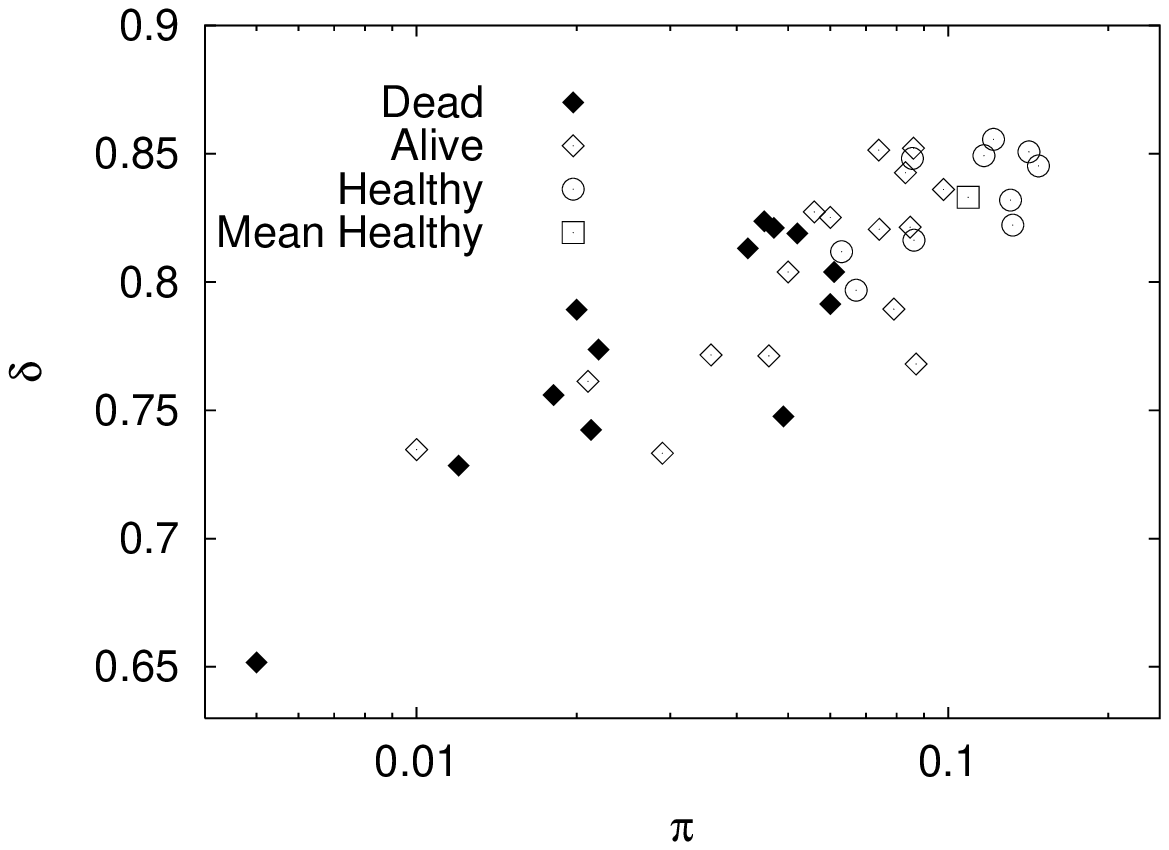}
\caption{\label{fig2} Positions of healthy (circles) and CHF (diamonds)
subjects in the physiological plane. The white diamonds
correspond to patients alive after the end of the experiment, while
the black ones to patients who were either dead or urgently transplanted.  
The white square (optimal condition, see text) represents the 
average position of healthy subjects in the plot.}
\end{center}
\end{figure}


\begin{references}
\bibitem{memorybeyondmemory} P. Allegrini, P. Grigolini, P. Hamilton,
L. Palatella, and G. Raffaelli, Phys. Rev. {\bf 65}, 041926 (2002).
\bibitem{vulpiani1} K. H. Andersen, P. Castiglione, A. Mazzino, and
A. Vulpiani, Eur. Phys. J. B {\bf 18}, 447 (2000).
\bibitem{vulpiani2} P. Castiglione, A. Mazzino,
P. Muratore-Ginnanneschi, A. Vulpiani, Physica D {\bf 134}, 75 (1999).
\bibitem{giacomo} P. Grigolini, L. Palatella, G. Raffaelli, Fractals,
{\bf 9}, 439 (2001).
\bibitem{allegro} P. Allegrini, P. Grigolini, B.J. West, Phys. Rev. E
{\bf 54}, 4760 (1996).
\bibitem{note} It is possible to prove that moving from the
dichotomous condition of Ref.\cite{allegro} to the many-values
condition of Refs. \cite{vulpiani1,vulpiani2} does not affect the
essential properties of the resulting anomalous diffusion. For the
sake of brevity, we do not show here this demonstration. Nevertheless,
the arguments of this letter rests on the dichotomous picture of
Ref. \cite{allegro}.
\bibitem{nicola} N. Scafetta, P. Hamilton, P. Grigolini, Fractals,{\bf
9}, 193 (2001).
\bibitem{note2} The authors of the earlier publication of
Ref. \cite{memorybeyondmemory} obtained a clear division into two
distinct groups. The counterpart of the physiological plane of this
letter is not defined with the same accuracy as that here
used. Nevertheless, the main reason for the existence of an overlap,
in an apparent conflict with the neat division into two groups of
Ref.\cite{memorybeyondmemory} is due to a different prescription to
deal with the atrial extrasystole.
\bibitem{test} M. Hollander, D. A. Wolfe, \emph{Nonparametric
Statistical Methods}, John Wiley and Sons, New York (1973).
\bibitem{stanley1} P.Ch. Ivanov, L.A. Nunes Amaral, A.L. Goldberger,
S. Havlin, M. G. Rosenblum, Z.R. Struzik, and H. E. Stanley, Nature
(London) {\bf 399}, 461 (1999).
\bibitem{stanley2} Y. Ashkenazy, P.Ch. Ivanov, S. Havlin, C. -K. Peng, A.L. Goldberger, and H.E. Stanley, Phys. Rev. Lett. {\bf 86}, 1900 (2001).
\end{references}
\end{document}